\documentclass[preprint,12pt]{elsarticle}

\usepackage{epsfig}
\usepackage{amssymb}
\usepackage{amsmath}
\usepackage{bm}
\usepackage{caption}
\journal{Journal of Magnetism and Magnetic Materials}

\begin{document}

\begin{frontmatter}



\title{How experimentally to detect  a solitary superconductivity in dirty ferromagnet-superconductor trilayers?}


\author{Maxim V. Avdeev}\ead{avdeev.maxim.kfu@gmail.com}
\author{Yurii N. Proshin}

\address{Kazan Federal University, Russian, Kazan, Kremlyovskaya 18, 4200008}

\begin{abstract}
We theoretically study the proximity effect in the
thin-film layered ferromagnet (F) - superconductor (S) heterostructures in F$_1$F$_2$S design.
We consider the boundary value problem for the Usadel-like equations in the case of so-called ``dirty'' limit.
The ``latent'' superconducting pairing interaction in F layers taken into account.
The focus is on the recipe of experimental preparation the state with so-called solitary superconductivity.    
We also propose and discuss the model of the superconducting spin valve based on F$_1$F$_2$S trilayers in solitary superconductivity regime.   
\end{abstract}


\end{frontmatter}


\section{Introduction}
\label{Intro}
The coexistence of two antagonistic phenomena -- superconductivity (S) and ferromagnetism (F)  in artificial layered structures is possible due to the proximity effect~\cite{deGennes_RMP_1964}.
The singlet superconducting condensate can penetrate from the S layer into the F layer,   
and nonmonotonically decays on very short distance about $\xi_I = \sqrt{D/2I}$ (where $I$ is
the exchange field and $D$ is the diffusion coefficient in a ferromagnet)  into the F layer. For strong ferromagnets such as Fe, Ni or Co the decay depth is approximately few nanometers. 
The peculiarity of the FS proximity effect and interplay  between the S and F parameter orders give rise to a number of
interesting phenomena and effects (see reviews~\cite{Izumov_UFN_2002, Buzdin_RMP_2005, Bergeret_RMP_2005} and the references therein), for example, the reentrant~\cite{Proshin_PRB_1997, Proshin_jetp_1998, Proshin_JLTP_2015} and solitary superconductivity~\cite{Proshin_JMMM_2015,Avdeev_jetplett_2015}, nonmonotonic behaviors of the critical temperature $T_c$ as function of the mutual alignment of the magnetizations of the F layers~\cite{Fominov_jetplett_2003, Fominov_jetplett_2010} and so on. 

Rich physics of the FS proximity effect and rapid progress in area of the spintronics and superconducting electronics make this field promising for spin valve applications. For example,
spin valve devices based on three-layer FS systems switched
by a weak external magnetic field were proposed in~\cite{Oh_APL_1997, Tagirov_PRL_1999, Buzdin_EPL_1999}. 
In a recent experimental work~\cite{Leksin_PRB_2015} the difference $\Delta T(\alpha) = T_c(\alpha) - T_c(0^\circ)$ (where $\alpha$ is angle between magnetizations of the adjacent F layers) was measured in F$_1$F$_2$S (Fe/Cu/Fe/Cu/Pb) trilayer design. 
It has been shown that its highest value reached at the perpendicular magnetic alignment when $\Delta T(\alpha = 90^\circ)\approx 40$\,mK. Further, the difference $\Delta T(\alpha = 90^\circ)\approx800$\,mK was achieved for a similar F$_1$F$_2$S (CrO$_2$/Cu/Ni/MoGe) system~\cite{Singh_PRX_2015}.  

Earlier we theoretically investigate a solitary superconductivity for F$_1$F$_2$S system~\cite{Proshin_JLTP_2015,Proshin_JMMM_2015,Avdeev_jetplett_2015}.
The solitary superconductivity corresponds to a localized region on the phase diagram
of $T_c(d_f)$, in which $T_c > 0$ and thickness $d_f$ belongs to region [$d_f^{*}$,$d_f^{**}$], where
$d_f^{*}> 0$. This occurs only at the antiparallel (AP) mutual magnetic aligned of F$_1$ and F$_2$ layers.
The superconductivity does not occur at the parallel (P) orientation that
makes relevant the study of states with
solitary superconductivity, as they may
prove to be the most promising for the
superconducting spin valve applications.

In main goal this work is how experimentally to detect the solitary superconductivity.

\section{Theoretical background}
\label{Theory}
Near the superconducting transition the
self-consistent equations for the
superconducting order parameters has the form~\cite{deGennes_RMP_1964}
\begin{equation}
\label{eq:self_consist}
\Delta_{i}(\mathbf{r})(\ln t + \ln\frac{T_{cs}}{T_{i}})=2\pi T_c\mathrm{Re}\sum_{\omega>0}^{\infty}\left( F_i(\mathbf{r},\omega)-\frac{\Delta_i(\mathbf{r})}{\omega}\right),\,\, i = (f1, f2, s),
\end{equation}
where $t=T_{\mathrm{c}}/T_\mathrm{cs}$ is the reduced critical temperature ($T_\mathrm{cs}$ and $T_i$ is the superconducting critical temperature for the bulk material (S and F$_i$, respectively) without spin exchange interaction), $\omega$ is the Matsubara frequency.

The pair amplitudes $F_{s,(i)}$ satisfy the Usadel-like equations~\cite{Uzadel_PRL_1970}
\begin{equation}
\begin{aligned}
\left[|\omega|-i I_{i}-\frac{D_{i}}{2}\dfrac{d^2}{dx^2}  \right]F_{i}(x,\omega)=\Delta_{i}(x).
\label{eq:pair_ampl_f}
\end{aligned}
\end{equation}
Here  $I_i$ is the exchange interaction in F layers, $D_{i}$ is the diffusion constant.

For pair amplitude $F_i$, we have Kupriyanov-Likichev like boundary conditions~\cite{Kupriyanov_jetp_1988} derived by microscopic approach in the work~\cite{Proshin_jetp_1998}.
For the F$_2$S and F$_1$F$_2$ interfaces they have the form
\begin{equation}
\begin{aligned}
\frac{4 D_s}{\sigma_s \upsilon_F^s}\dfrac{d}{dx} F_{s} =\frac{4 D_{f2}}{\sigma_{f2} \upsilon_F^{f2}}\dfrac{d}{dx} F_{f2} =F_s-F_{f2},\\
\frac{4 D_{f1}}{\sigma_{f1} \upsilon_F^{f1}}\dfrac{d}{dx} F_{f1} =\frac{4 D_{f2}}{\sigma_{f2} \upsilon_F^{f2}}\dfrac{d}{dx} F_{f2} =F_{f2}-F_{f1}.
\label{eq:boundary_cond}
\end{aligned}
\end{equation}
The boundary conditions at the outer surfaces have the form
\begin{equation}
\label{eq:free_boundary_cond}
\dfrac{d}{dx} F_{s,f}=0.
\end{equation}
The parameters  $\sigma_{s}$ and $\sigma_f$ are the transparencies from the S and F side, respectively~\cite{Izumov_UFN_2002} and $\upsilon_F^i$ is Fermi velocity.

Then, the solutions of eqs.~(\ref{eq:pair_ampl_f}) for F$_1$F$_2$S trilayers have the form
\begin{equation}
\begin{aligned}
\label{eq:sol_ffs}
F_1 = \dfrac{\Delta_1}{\omega - iI_1} + C_1(\omega)\cosh k_{I1} (x + d_{f1} + d_{f2}),\\%\, (-d_{f1}-d_{f2} < x < -d_{f2});\\
F_2 = \dfrac{\Delta_2}{\omega - iI_2} + A(\omega)\cosh k_{I2}x + B(\omega)\sinh k_{I2}x,\\%\, (-d_{f2} < x < 0);\\
F_s = \dfrac{\Delta_s}{\omega} + C_s(\omega)\cosh k_s(x-d_s),\\%\, (0 < x < d_s);
\end{aligned}
\end{equation}
where $k_s^2 = {2\omega}/{D_s}$, $k_I^2 = 2({\omega - iI})/{D_f}$.
The set of solutions~(\ref{eq:sol_ffs}) and the appropriate boundary conditions (\ref{eq:boundary_cond}), (\ref{eq:free_boundary_cond}) are sufficient to determine the coefficients $C_1$, $A$, $B$, $C_s$ that are linear combinations of the gaps $\Delta_s$, $\Delta_1$, $\Delta_2$.
Finally, inserting the solutions (\ref{eq:sol_ffs}) into equations (\ref{eq:self_consist}) and solving the resulting secular equation, we calculate the critical temperature $T_\mathrm{c}$ for the F$_1$F$_2$S system.

\section{Results and discussion}\label{Results}
In this section, we focus on preparation of the F$_1$F$_2$S system model to detect the solitary superconductivity. We also discuss its of the  spin valve applications. 
As was mentioned above the solitary superconductivity in  FFS system may be observed only in the AP state~\cite{Proshin_JLTP_2015, Proshin_JMMM_2015, Avdeev_jetplett_2015}. 
%Without loss of generality we assume that outer F layer  
In order to prepare state with a well-defined solitary superconductivity 
experimentalists should adhere to the following algorithm. Firstly the superconductivity in the FS bilayer is investigated. Without loss of generality the F layer thickness is assumed much greater than coherence length $d_f \gg \xi_I$. Further at the fixed $d_f$ the S layer thickness $d_s$ should be changed to find the critical value $d_s^{*}$ at which the critical temperature $T_c$ is close to zero. 
\begin{figure}[ht!]
	\includegraphics[width=\linewidth]{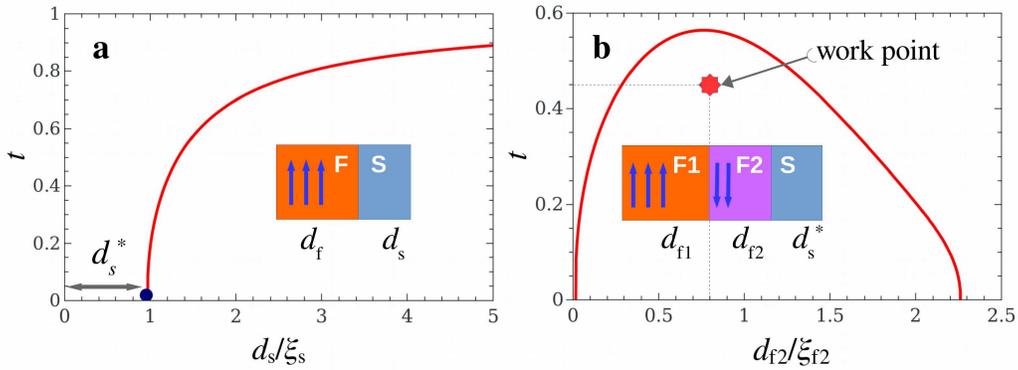}
	\caption{\label{spin_valve}(Color online) (a) The reduced critical temperature versus the S layer thickness for the FS bilayer system ($d_s^{*} = 0.95\xi_s$). (b) The phase diagrams of the solitary superconductivity for the FS trilayer at AP state. The red star denote the work point of spin valve (see text). 
		Other parameters of the system are: $l_s/\xi_{s}=0.5$, $\sigma_s=1$, $\sigma_{f2}=3$, $\sigma_{f1}=1$, $d_{f1}/\xi_{I1}=10$, $l_{f}/\xi_{I}=0.5$,  $I/\pi T_{cs}=10$,  $T_{cs}/T_{cf}=1$.}
\end{figure}
Thus, in figure~\ref{spin_valve}a the dependence of the reduced critical temperature $t$ on the S layer thickness $d_s$ for the FS bilayer system is shown. It is seen that superconductivity disappears at $d_s^{*} \approx \xi_s$ ( 
for convenience, the thickness $d_s$ of the S layer and mean free path $l_s$ is normalized on the coherence length $\xi_s = \sqrt{D_s/2\pi T_{cs}}$, while all length relating to the both F$_1$ and F$_2$ layers are normalized on the coherence lengths $\xi_{I1,2} = \sqrt{D_{f1,2}/2I_{1,2}}$ respectively). 

Next, the samples of the F$_1$F$_2$S trilayers should be prepared with various thicknesses $d_{f2}$ of the intermediate  F$_2$ layer at fixed both $d_{f1}$ and $d_s^{*}$ thicknesses. It is important that the F$_1$F$_2$S system should be in the AP state. In figure~\ref{spin_valve}b it is clearly seen that a superconductivity is restored with the $d_{f2}$ increase.  
The critical temperature $T_c$ begins to rise and reaches to the maximum value $t = 0.565$ at $d_{f2} \approx 0.76\xi_{I2}$.
With further increase of the intermediate layer thickness, the critical temperature decreases monotonically up to zero.  As was mentioned above, such an extraordinary nonmonotonic dependence $T_c(d_{f2})$ is called solitary superconductivity. Note again that for the parallel state, the system is always in the resistive state~\cite{Proshin_JLTP_2015, Proshin_JMMM_2015, Avdeev_jetplett_2015} and hence, the difference $\Delta T = T_c^{AP} - T_c^P\equiv T_c^{AP}$ is a maximal. The physical reason for this phenomenon is simple: the exchange field of the intermediate F$_2$ layer, which is a strong depairing factor for the superconducting state,  is partially compensated at AP state, herewith the ``effective'' exchange field becomes smaller and superconductivity appears. 
For more complete compensations of the effective exchange field the soft magnetic materials with low Curie temperature (such as Ni$_{1-x}$Cu$_x$ or Pr) alloys should be used for both F layers~\cite{Proshin_JLTP_2015, Proshin_JMMM_2015, Avdeev_jetplett_2015}.

The deviation of the magnetic alignment from the AP state results in the fast vanishing of this compensation and, as a result, in the reduction of the critical temperature~\cite{Proshin_JLTP_2015, Proshin_JMMM_2015, Avdeev_jetplett_2015}.  This important feature of the system with solitary superconductivity makes them the most promising for superconducting spin valve applications, because the high values of difference $\Delta T$ are necessary  for stable operation of spin valve. In order to consider the model of spin valve we choose the work point in the phase diagram at $d_{f2} = 0.8\xi_{I2}$ and $T^{*} = 0.45T_{cs}$ (the red octagonal star in figure~\ref{spin_valve}b). At the AP state $T^{*} < T_c$ and system is in the superconducting state. Upon a change in the magnetic alignment from the AP state to the P state by weak external magnetic field, the superconductivity disappears, and system switches in resistive state.   
   
\section{Summary}
In this work we consider the feature of states with a solitary superconductivity for the asymmetrical F$_1$F$_2$S trilayers. 
The appearance of these states for dirty F$_1$F$_2$S trilayers is a result of partial compensation of \emph{effective} exchange fields of F layers at AP state.
Based on out results  we propose the way for observation of solitary superconductivity in real experiments.
We also discus the spin valve model in solitary superconductivity regime.  We shown that in this case  the  difference $\Delta T$ between two superconducting and resistive states is maximal.  It leads to more stable operation of spin valve.

\section*{Acknowledgement}
The work is partially supported by the RFBR (16-02-01016). YNP is also thankful to the subsidy allocated to Kazan Federal University for performing the project part of the state assignment in the area of scientific activities.



  \bibliographystyle{elsarticle-num}
  \bibliography{bibdatabase}


%\begin{thebibliography}{00}


%\bibitem{}

%\end{thebibliography}
\end{document}